\pdfoutput=1
\documentclass{elsart1p}
\usepackage{graphicx,amssymb,lineno}
\usepackage{amstext,amssymb,amsmath,amscd}
\usepackage{natbib}

\makeatletter
\def\elsartstyle{%
    \def\normalsize{\@setfontsize\normalsize\@xiipt{14.5}}
    \def\small{\@setfontsize\small\@xipt{13.6}}
    \let\footnotesize=\small
    \def\large{\@setfontsize\large\@xivpt{18}}
    \def\Large{\@setfontsize\Large\@xviipt{22}}
    \skip\@mpfootins = 18\p@ \@plus 2\p@
    \normalsize
}
\@ifundefined{square}{}{\let\Box\square}
\makeatother

\begin{document}

\begin{frontmatter}
\title{CMB data analysis and sparsity}

\author[1]{Abrial P.}
\ead{pabrial@cea.fr}
\author[1]{Moudden Y.}
\author[1,2]{Starck J.-L.}
\author[3]{Fadili J.} 
\author[2]{Delabrouille J.}
\author[4]{Nguyen M.K.}
\address[1]{DAPNIA/SEDI-SAP, Service d'Astrophysique,
CEA/Saclay, 91191 Gif sur Yvette, France.}
\address[2]{Laboratoire APC, 11 place Marcelin Berthelot 75231 Paris Cedex 05, France.}
\address[3]{GREYC - CNRS UMR 6072/ENSICAEN,  14050 Caen Cedex, France.}
\address[4]{ETIS - CNRS UMR 8051/ENSEA/Universit\'e de Cergy-Pontoise, 95014 Cergy-Pontoise, France.}

\begin{abstract}
 The statistical analysis of the soon to come Planck satellite CMB data will help set tighter bounds on major cosmological parameters. On the way, a number of practical difficulties need to be tackled, notably that several other astrophysical sources emit radiation in the frequency range of CMB observations. Some level of residual contributions, most significantly in the galactic region and at the locations of strong radio point sources will unavoidably contaminate the estimated spherical CMB map. Masking out these regions is common practice but the gaps in the data need proper handling. In order to restore the \emph{stationarity} of a partly incomplete CMB map and thus lower the impact of the gaps on non-local statistical tests, we developed an inpainting algorithm on the sphere based on a sparse representation of the data,  to fill in and interpolate across the masked regions. 
\end{abstract}

\end{frontmatter}
\section{Introduction}
\label{sect:intro}
The analysis of the slight fluctuations in the Cosmic Microwave Background radiation field (CMB), for which evidence was found for the first time in the early 1990's in the observations made by COBE~\cite{astro:COBE2},  is a major issue in modern cosmology  as these are strongly related to  the cosmological scenarios describing the properties and evolution of our Universe. In the 'Big Bang' model, the map of CMB fluctuations is an imprint of primordial fluctuations in matter density from a time when the temperature of the Universe in quasi thermal equilibrium, was high enough above  3000~K for matter and radiation to be tightly coupled. With gravity, the density fluctuations collapsed into large scale structures such as galaxies or clusters of galaxies. Due to the expansion of the nearly transparent Universe in which they were set free, the CMB photons are now observed in the microwave range, while still distributed according to an almost perfect Black Body emission law. 
Full-sky multi-spectral observations of the CMB with unprecedented sensitivity and angular resolution are expected from the ESA's Planck
mission, which is to be launched in september 2008. The statistical analysis of this data set will help set tighter bounds on major cosmological parameters. On the way, there are a number of practical difficulties that need to be overcome and notably that several other astrophysical sources also emit radiation in the frequency range used for CMB observations~\cite{fb-rg99}. The task of separating the observed mixture maps back into the different contributing astrophysical components in order to isolate the CMB properly turns out to be a redoubtable and strenuous inverse problem  for which dedicated methods and algorithms are currently being actively designed (\emph{e.g.}~\cite{astro:2005MNRAS.364.1185P,starck:bobin06,wlens:pires06} and references therein). Some level of residual contributions, most significantly in the galactic region and at the locations of strong radio point sources will unavoidably contaminate the estimated spherical CMB maps. Therefore, it is common practice to mask out those parts of the data (\emph{e.g.} using the mask shown on figure~\ref{Figure:cmb_scale_wmap_inpainting}, provided by the WMAP\footnote{http://map.gsfc.nasa.gov} team) 
in order for instance to reliably assess the non-Gaussianity of the CMB field through estimated higher order statistics (\emph{e.g.} skewness, kurtosis ) in various representations (\emph{e.g. wavelet, curvelet, etc.}) \cite{starck:jin05} or to estimate the power spectrum of the CMB spatial fluctuations. But the gaps in the data thus created need to be handled properly.
In order to restore the \emph{stationarity} of a partly incomplete CMB map and thus drastically lower the impact of the missing patches on the estimated measures of non-gaussianity or on any other non-local statistical test, we developed an inpainting algorithm on the sphere to fill in and interpolate across the masked regions. The grounds for our inpainting scheme are in the notion of \emph{sparsity} of the representation of a data set as is quickly discussed in section~\ref{sect:one}. The proposed gap-filling algorithm is described in section~\ref{sect:two}. Several numerical experiments  were conducted  on synthetic data in the context of CMB data analysis and the results of these are reported in section~\ref{sect:three}.    
\section{Sparsity and CMB data maps}
\label{sect:one}
Consider a vector of observations $y \in\mathbb{R}^n$ the entries of which are the pixels of some spherical map, or the samples of some 1D signal or 2D image ec. in some other context. A common processing task is then to decompose the data $y$ into its elementary building blocks as in the following generative model :  
\begin{equation}
 y = \sum_{i} \alpha_i \phi_i + \eta
\end{equation}
that is a linear combination of relevant waveforms $\phi_i \in \mathbb{R}^n$  with weights $\alpha_i$. Here $\eta$ represents possible contamination by additive, typically Gaussian white noise.  Given data $y\in \mathbb{R}^n$, one then wants to recover the underlying structures that is to say estimate a set of waveforms $\phi_i$ that build the data and their corresponding weights $\tilde{\alpha}_i$. The solution to this estimation problem will depend heavily on the available prior information. We will assume here, for instance, that one is given \emph{a priori} a basis, a frame or a large redundant dictionary of waveforms from which to select a good subset. 
For data mapped to the sphere such as full-sky CMB maps, available invertible transforms include the spherical harmonics and several wavelet transforms. Software packages such as Healpix
\footnote {http://www.eso.org/science/healpix}
~\cite{pixel:healpix} or Glesp~\cite{pixel:glesp} provide approximate digital spherical harmonic transform routines based on their specific pixelization schemes.  Schr{\"o}der and  Sweldens~\cite{wave:sweldens95a} have developed an orthogonal wavelet  transform on the sphere; Freeden \emph{et al.} give a wavelet transform/reconstruction scheme on the sphere which is based on the spherical harmonic transform~\cite{freeden97}. Following this idea, Starck \emph{et al.}~\cite{starck:sta05_2}  proposed an invertible isotropic undecimated wavelet transform (UWT) on the sphere which preserves the same desirable properties as the standard isotropic UWT for flat 2D maps~\cite{starck:book98}.  Recently, building on the latter and the Healpix geometry, other multiscale transforms such as the pyramidal wavelet transform, the ridgelet transform and the curvelet transform were extended to handle spherical maps~\cite{starck:sta05_2}. 
\begin{figure}
\vbox{
\centerline{
\hbox{
\includegraphics[width = 7cm]{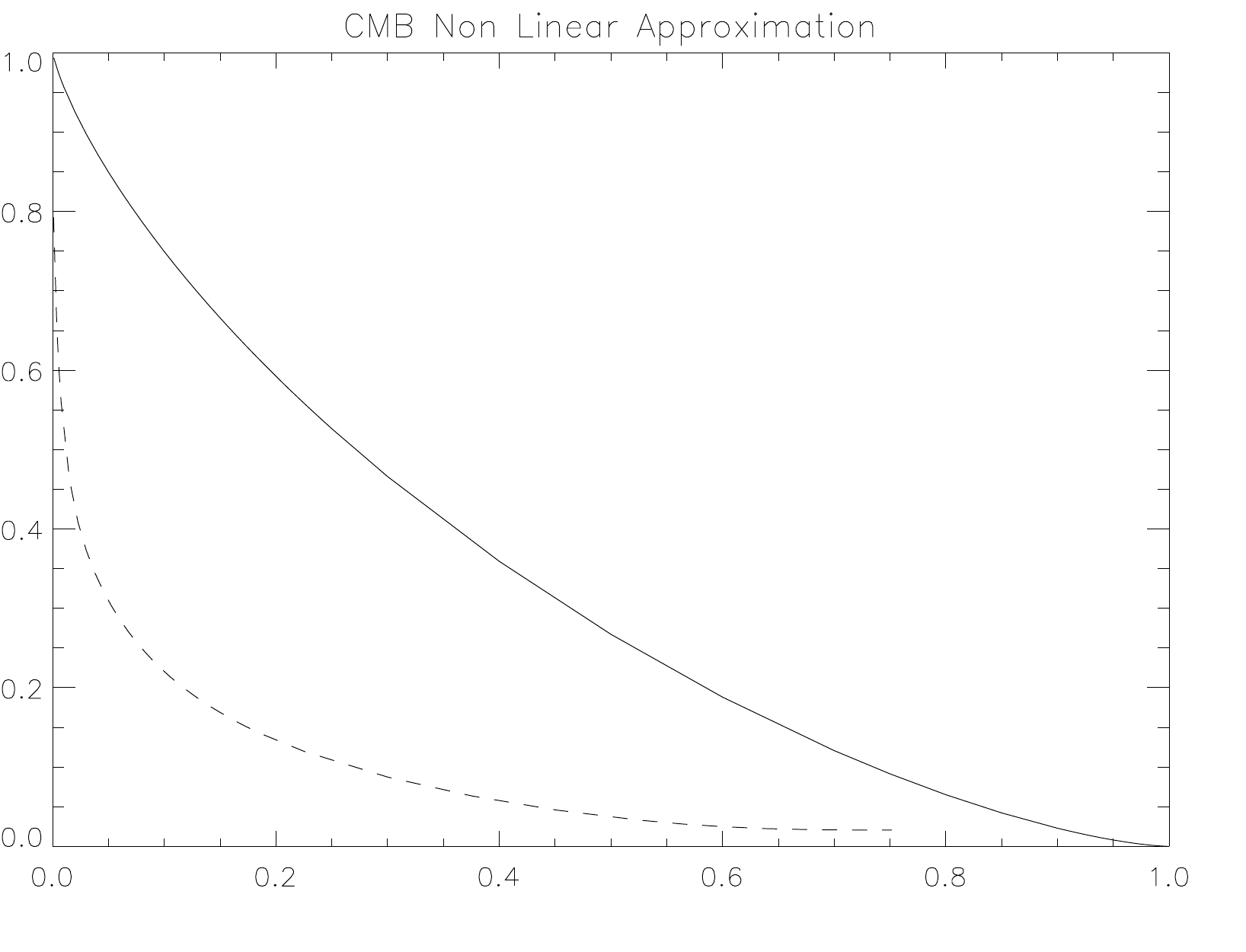}
}
}
}
\caption{   $\ell_2$ non-linear approximation relative error curve of a simulated CMB map in the spherical harmonics representation (dotted line) and in the \emph{pixel} representation (continuous line) as a function of the fraction of largest coefficients in the truncated reconstruction series. The CMB map appears to have a sparser representation in the spherical harmonics decomposition.}
\label{fig:nlapprox}
\end{figure}

Any $y \in\mathbb{R}^n$ obviously has an exact representation over any complete basis of $\in\mathbb{R}^n$ or several such exact representations in the case of redundant overcomplete dictionaries. However, these representations are not equally interesting in terms of data modeling or feature detection and there is a strong \emph{a priori} in favor of sparse representations of $y$ that use only a small number of waveforms : exhibiting a sparse representation of the data $y\in\mathbb{R}^n$ suddenly makes \emph{information} more concise and possibly more interpretable. Still, building sparse representations or approximations of structured data by solving
\begin{equation}
 \min_{\alpha}  \|\alpha\|_{{ \ell_0}} \mbox{ subject to } y = \Phi \alpha
\label{eqn_l0}
\end{equation}
is about selecting the smallest subset of waveforms from a possibly redundant dictionary $\Phi$,  that will linearly combine to reproduce the salient features of a given signal or image $y$, and in general this is a hard combinatorial problem. A number of algorithms have been proposed in an attempt to solve~(\ref{eqn_l0}) directly or \emph{relaxed} or approximate versions of this problem \emph{e.g.} 
\begin{equation}
\textrm{BP : } \quad \min_{\alpha}  \|\alpha\|_{{ \ell_1}} \mbox{ subject to } y = \Phi \alpha
\label{eqn_bp}
\end{equation}\begin{equation}
\textrm{BPDN : } \quad \min_{\alpha}  \frac{1}{2} \| y -  \Phi \alpha \|^2_{{ \ell_2}} + \lambda \|\alpha\|_{{ \ell_1}} \quad \lambda > 0
\label{eqn_bpdn}
\end{equation}
where sparsity is measured using  ${\ell_1}$  norm in place of the ${\ell_0}$ counting norm. This large set of  algorithms includes the greedy Matching Pursuit (MP)~\cite{wave:mallat93}, Basis Pursuit (BP) and Basis Pursuit Denoising (BPDN)~\cite{wave:donoho98},  LARS~\cite{mca:lars}, Stomp~\cite{donoho:stgomp}, MCA~\cite{mca:sta04} and Polytope Faces Pursuit~\cite{mca:PFP}. A gradient descent algorithm to solve ~(\ref{eqn_bpdn}) is discribed in ~\cite{alliney1,alliney2}. The conditions under which each of these algorithms provides a unique sparse solution and the conditions under which these solutions actually coincide with the optimal solution to problem~(\ref{eqn_l0}) have been recently explored by a number of authors(\emph{e.g.}~\cite{Donoho-Elad,cur:elad02,miki:fuchs}. They showed that the proposed methods are able to recover the sparsest solution provided this solution is indeed sparse enough and the dictionary is sufficiently incoherent.\\
\begin{figure}
\vbox{
\centerline{
\hbox{
\includegraphics[height = 4.6cm,width = 6.8cm]{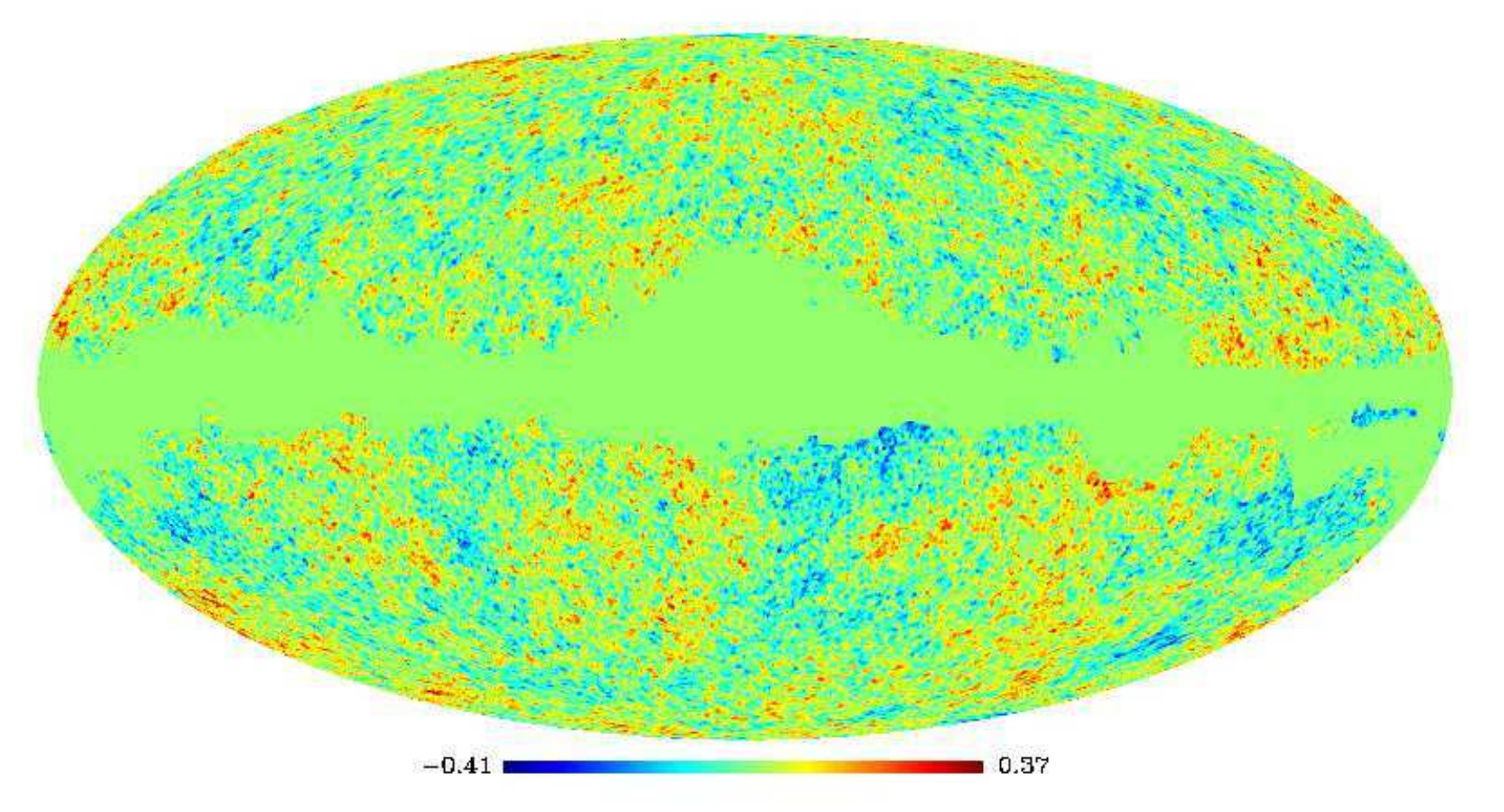}
\includegraphics[height = 4.6cm, width = 6.8cm]{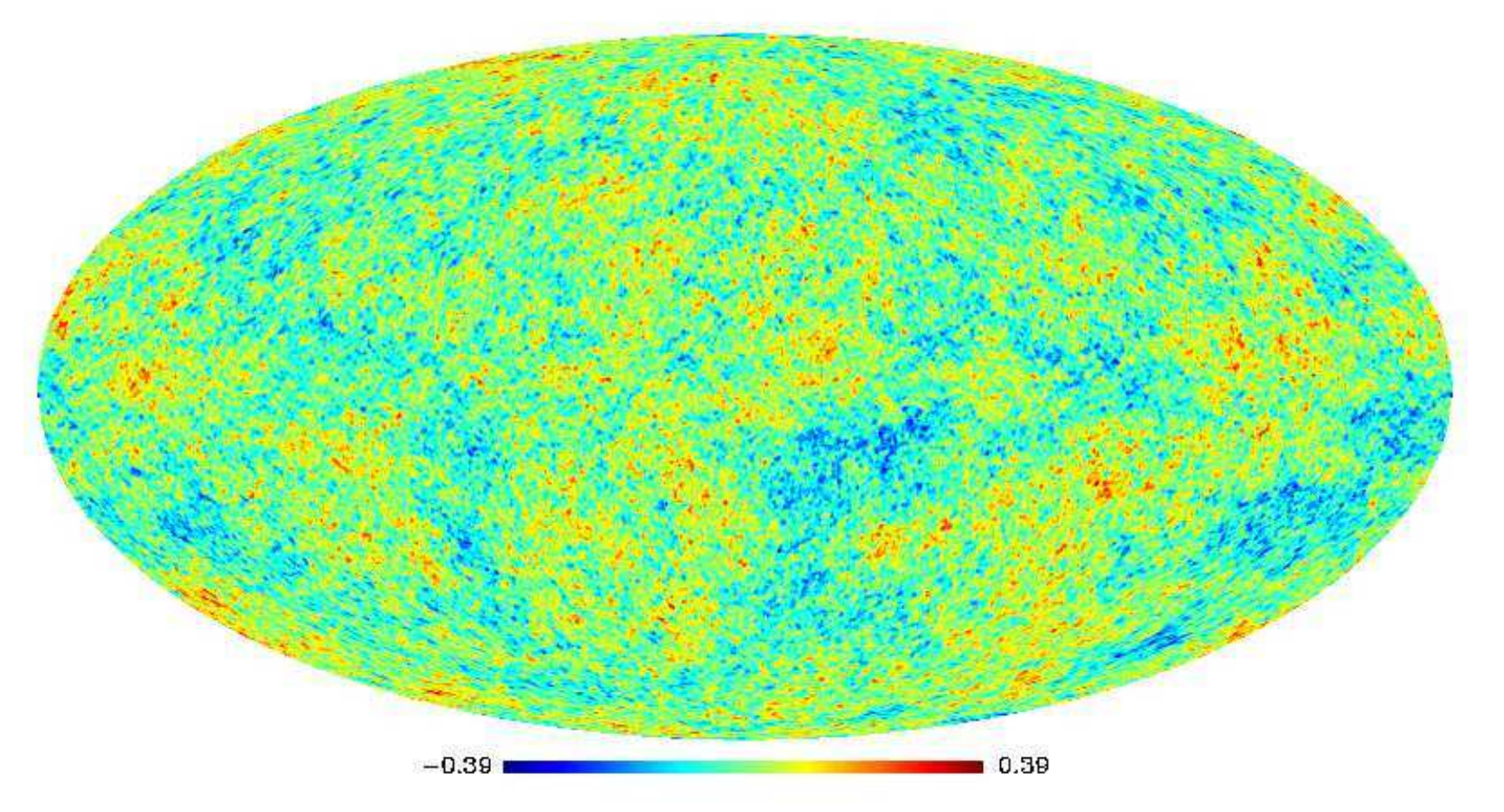}
}
}
}
\caption{\textbf{left :} WMAP data. \textbf{right :} WMAP inpainted map.}
\label{Figure:cmb_wmap_inpainting}
\end{figure}

\indent
CMB data analysis is a challenging application for sparse signal processing methods. As mentioned previously, CMB is well modeled as a mixture of different astrophysical contributions which may have extremely sparse representation in different bases. For instance the map of Sunyaev-Zeldovich galaxy clusters is well represented in an undecimated isotropic wavelet dictionary while the structured galactic dust clouds or the arcs in the map of synchrotron radiation are sparsely represented using a dictionary of anisotropic waveforms such as curvelets. Separating the different components can be formulated as a problem of building a sparse representation for multichannel data in a large dictionary of morpho-spectral waveforms. This was proposed in~\cite{starck:bobin06}  and encouraging results are reported in~\cite{mca:bobin_ada4}.  Our concern here is with correctly handling the gaps in the separated CMB map and interestingly this issue can also be approached on the side of sparse data processing. Indeed, although the CMB proper is well modeled as a realization of a stationary Gaussian random field on the sphere, it is not deprived of structure and so the concept of sparsity is still valid. The power spectrum of the spatial correlations in the CMB is not flat and the complex $a_{\ell, m}$ with $0\ge m\ge\ell$ of the CMB are well modeled as independent
 Gaussian variables with variance $C_\ell$ rapidly decreasing  towards zero as  the multipole number $\ell$ grows. Hence, the empirical distribution of all these coefficients together
is sparse. 
A convenient way to exhibit the sparsity of a specific representation of  $y$ is to look at the $\ell_2$ non-linear approximation relative error curve \emph{i.e.} at the fractional reconstruction error  obtained as a function of $N$ when a truncated reconstruction series, where only the $N$ largest coefficients are kept, is used as an approximation to the initial $y$. The graph on figure~\ref{fig:nlapprox} shows that the CMB has a sparse representation in the spherical harmonics decomposition. This important feature of the CMB radiation field is what is strongly relied on in order to fill in the gaps in an incomplete CMB map using the inpainting algorithm derived in the next section. 
\section{Full-sky CMB inpainting  based on sparsity }
\label{sect:two}
Inpainting refers to a set of methods used to alter images in a way that is undetectable to people who are unaware of the original images. There are numerous motivations for such tools among which removing scratches or objects  in digitized photographs, removing overlayed text or graphics,  filling-in missing blocks in unreliably transmitted images, predicting values in images for better compression or image upsampling. Inpainting algorithms strive to interpolate through the gaps in the image relying on the available pixels, the continuation of edges, the periodicity of textures, etc.  The preservation of edges and texture, in other words discontinuities, across gaps has attracted much interest, and many contributions have been proposed to solve this interpolation task. Non-texture image inpainting has received considerable interest and excitement since the pioneering paper by Masnou and Morel~\cite{Masnou02} who proposed variational principles for image disocclusion. A recent wave of interest in inpainting has started from the recent contributions of Sapiro \emph{et al.}~\cite{text:sapiro3}, followed by Chan and Shen~\cite{text:chan1}. In these works, authors point to the importance of geometry and design anisotropic diffusion PDEs to fill in gaps by smooth continuation of isophotes. PDE methods have been shown to perform well on piecewise smooth functions.

Another inpainting algorithm is the one described in~\cite{starck:elad05}. This method relies strongly on the ideas reviewed in the previous section of sparsity and the construction of sparse representations in large redundant dictionaries, thus promoting a rather different approach of the inpainting problem. To make the link between building sparse representations and inpainting, consider for instance the effect of a rectangular gap on the set of Fourier coefficients of a  monochromatic sinewave : in the Fourier domain, the consequence is a spread of the spectral lines of the initial sinewave due to convolution by the Fourier transform of the gap function. Also, due to the non-locality of the Fourier basis functions it takes a large number of coefficients to account for the inserted gap, which is something known as the  Gibbs effect. The proposed inpainting algorithm would deal with the present deconvolution problem in the spectral domain by, very schematically, solving a detection problem :  the complete monochromatic sinewave is recovered by iteratively selecting the largest coefficients in the Fourier domain with which a sparse signal is built which exactly fits the data outside the gap with no attempt to actually fit the zeros replacing missing data in the gap.

Extending the method in~\cite{starck:elad05} to handle incomplete maps on the sphere is straightforward. Consider a discrete spherical data map $y\in\mathbb{R}^n$ and a diagonal $n \times n$ matrix  $M$ such that ones on  the diagonal of $M$ indicate that the corresponding pixels in $y$ are valid data while zeros indicate invalid data. With a slight modification of the BP objective function~(\ref{eqn_bp}) as follows : 
\begin{equation}\label{inp:model}
\textrm{BP : } \quad \min_{\alpha}  \|\alpha\|_{{ \ell_1}} \mbox{ subject to } M\Phi \alpha =  M y
\end{equation}
we are preventing the sparse model under construction from attempting to fit the invalid data. We are seeking a solution $\alpha$ to a linear system of equations  with minimum $\ell_1$ norm. It appears clearly that features in the original data $y$ which are associated with atoms in $\Phi$ that are orthogonal to the pixel subspace defined by $M$ are lost and cannot be recovered with the described method. Different algorithms referred to in the previous section can be used to solve this minimization problem.  We propose that a satisfactory solution can be reached using an iterative thresholding process as in~\cite{mca:sta04} and~\cite{donoho:stgomp}. The algorithm in~\cite{mca:sta04} is simply modified so that the full residual is multiplied by $M$ after each residual estimation :
\begin{center}
\begin{minipage}[b]{0.9\linewidth}
\vspace{0.05in}
\footnotesize{\textsf{1. Set the number of iterations $I_{\max}$, the initial threshold $\lambda^{(0)} $ and $\tilde{y}^{(0)} = 0$.\\}

\textsf{2. While  $\lambda^{(t)}$ is greater than a given lower bound $\lambda_{\min}$ (e.g. may depend on the noise standard deviation), proceed with the following iteration :\\}

\hspace{0.15in} \textsf{--  Compute the residual term : $r^{(t)} = y - \tilde{y}^{(t-1)}$   }

\hspace{0.15in} \textsf{-- Thresholding :  $  \alpha^{(t)} = \delta_{\lambda^{(t)}}\left(  \Phi^{-1} \left( Mr^{(t)}+ \tilde{y}^{(t-1)}\right) \right)$ }

\hspace{0.15in} \textsf{-- Reconstruction : $\tilde{y}^{(t)} = \Phi \alpha^{(t)} $}

\hspace{0.15in} \textsf{-- Decrease the threshold $\lambda^{(t+1)}< \lambda^{(t)}$ following a given strategy.}
}
\vspace{0.05in}
\end{minipage}
\end{center} 
The way the threshold is decreased at each step is an important feature of the algorithm which will determine the speed and the quality of the inpainting. Refer to~\cite{starck:bobin06_tip} for suggestions and comparisons of different strategies. Also, current work is on improving the stability of our algorithm by enforcing additional constraints on the reconstructed map. For instance  a total variation penalty is shown in~\cite{starck:elad05} to enhance the recovery of piece-wise smooth components. Here, we ask instead for the regularity across the gaps of some localized statistics. Preliminary results show an improved stability when we impose that the empirical variance of the coefficients on each scale of an undecimated wavelet packet decomposition of the sparse inpainted map be \emph{equal} outside and inside the masked areas. These constraints are enforced  at each step $t$ of the iterative algorithm given above on the current inpainted map $\Phi \alpha^{(t)}$. The right choice of an undecimated wavelet packet decomposition that is of a partition of the multipole $\ell$ domain is another tunable feature of the algorithm which we are working on. It appears that a suitable partition should achieve a compromise between the need for bands to be large enough for the wavelet packet coefficients to be well localized in or out of the mask on the sphere and  the need for a fine partition in order to correctly adjust to our prior knowledge of the CMB power spectrum. 
\begin{figure}
\vbox{
\centerline{
\hbox{
\includegraphics[height=9cm,width=13cm,angle=180]{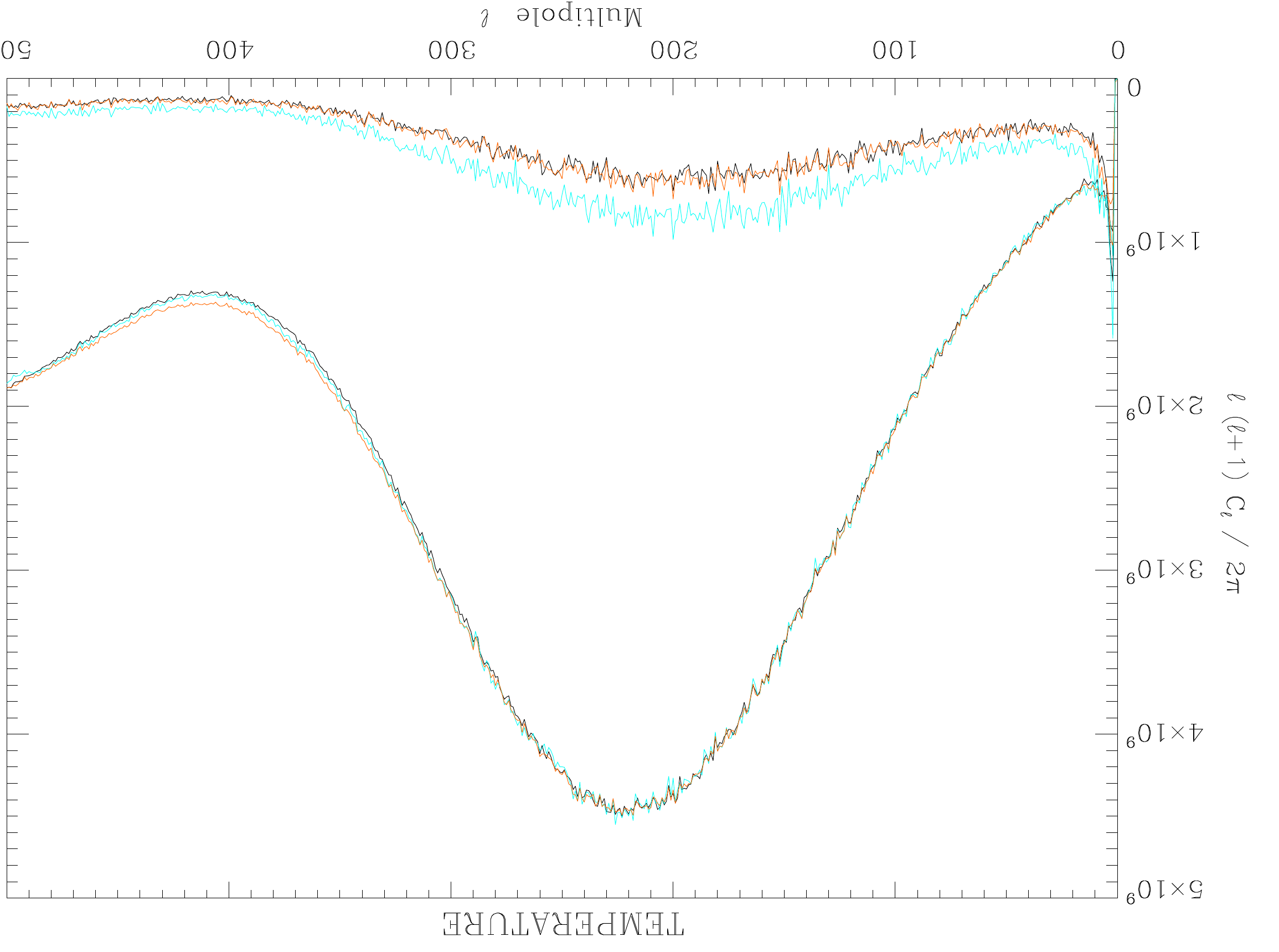}}
}
}
\caption{ Power spectrum of the CMB estimated from complete simulated CMB maps (black), from masked maps corrected for sky coverage (orange), inpainted maps (blue). The lower curves give the empirical variance (multiplied by 2 for visual purposes) as a function of $\ell$ of the three estimators of the power spectrum.}
\label{Figure:powspec}
\end{figure}
\section{Numerical experiments}
\label{sect:three}
\begin{figure}
\vbox{
\centerline{
\hbox{
\includegraphics[height=4.5cm,width=6.5cm]{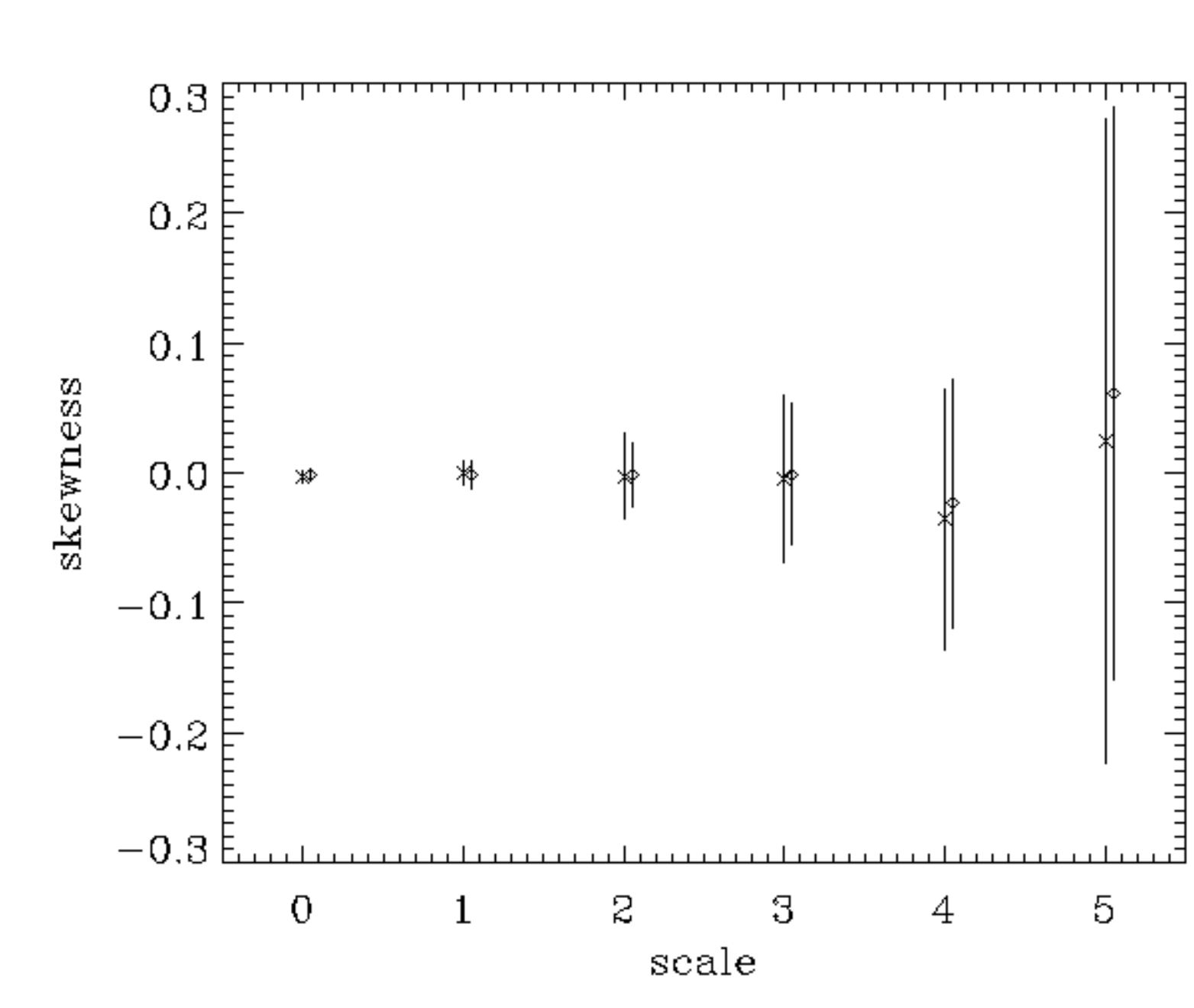}
\includegraphics[height=4.5cm,width=6.5cm]{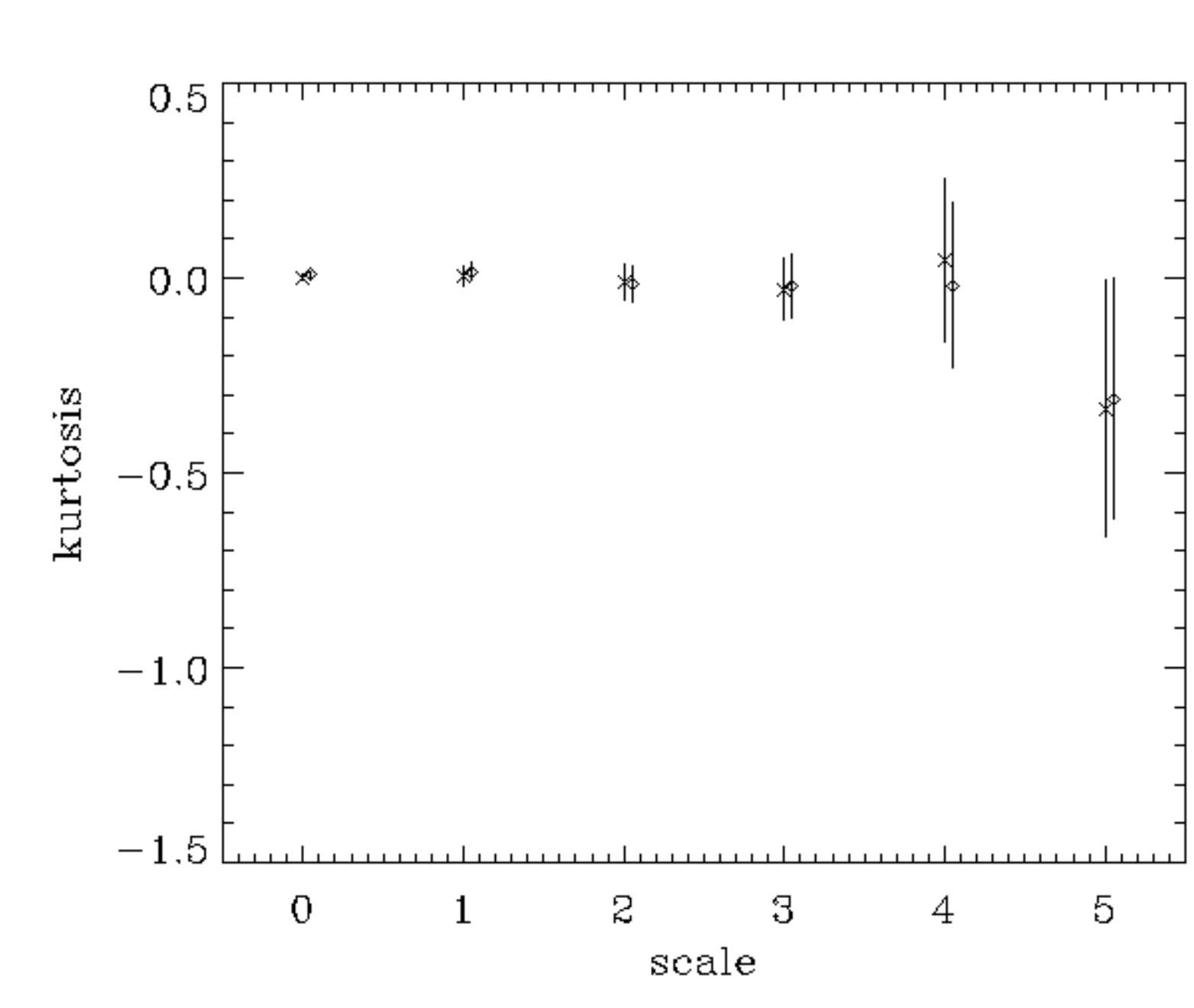}
}
}
}
\caption{ Horizontally is the scale number increasing for lower frequencies. \textbf{Left :} skewness of the wavelet coefficients at a given scale of the original complete spherical CMB map (o) and of the inpainted map (x). \textbf{Right :} kurtosis of the wavelet coefficients at a given scale of the original complete spherical CMB map (o) and of the inpainted map (x). RMS error bars were estimated on a small set of fifteen simulated complete CMB maps.}
\label{Figure:mca_skew_kur}
\end{figure}
\begin{figure}
\vbox{
\centerline{
\hbox{
\includegraphics[height=4.5cm,width=6.8cm]{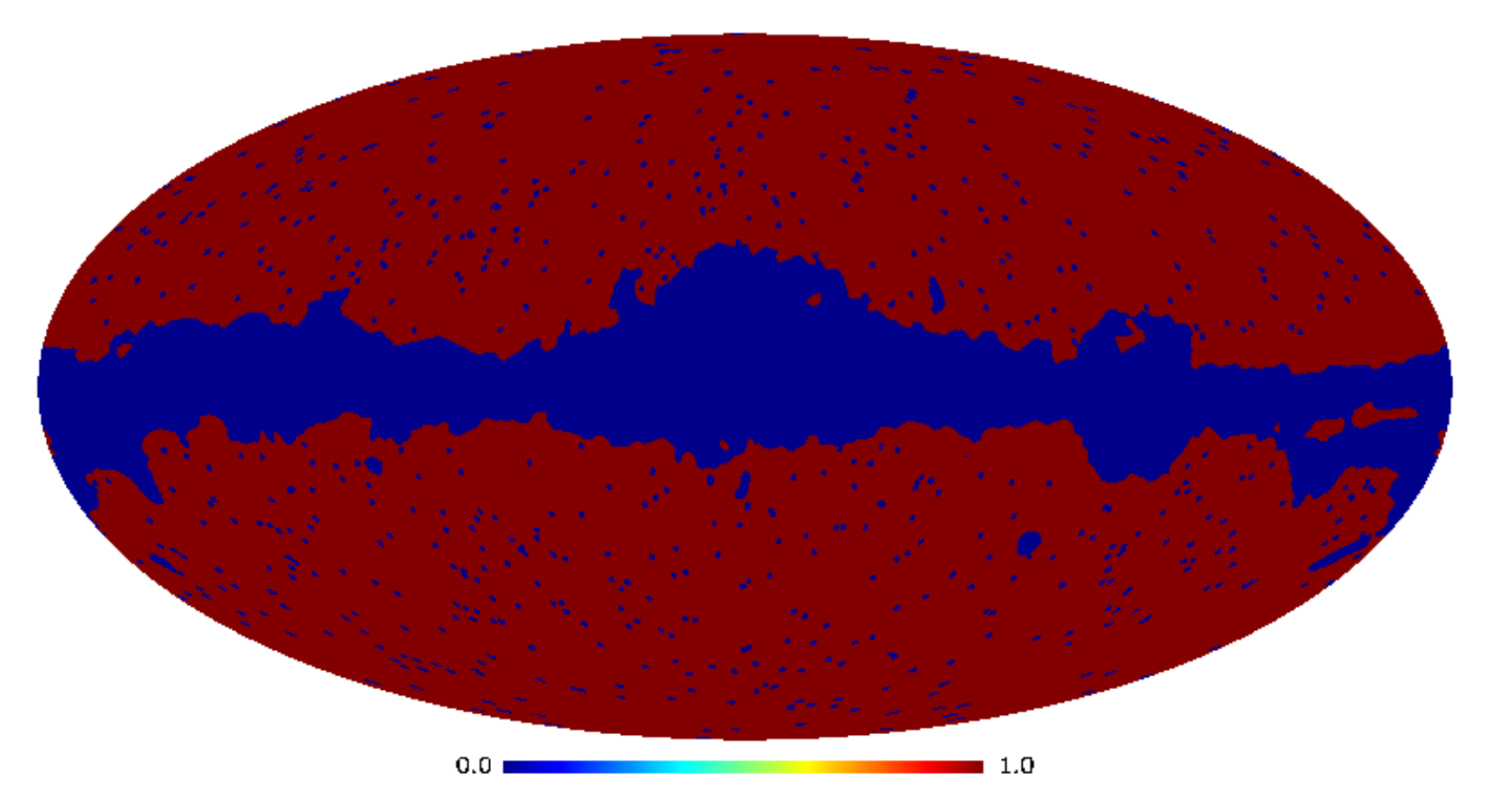}
\includegraphics[height=4.5cm,width=6.8cm]{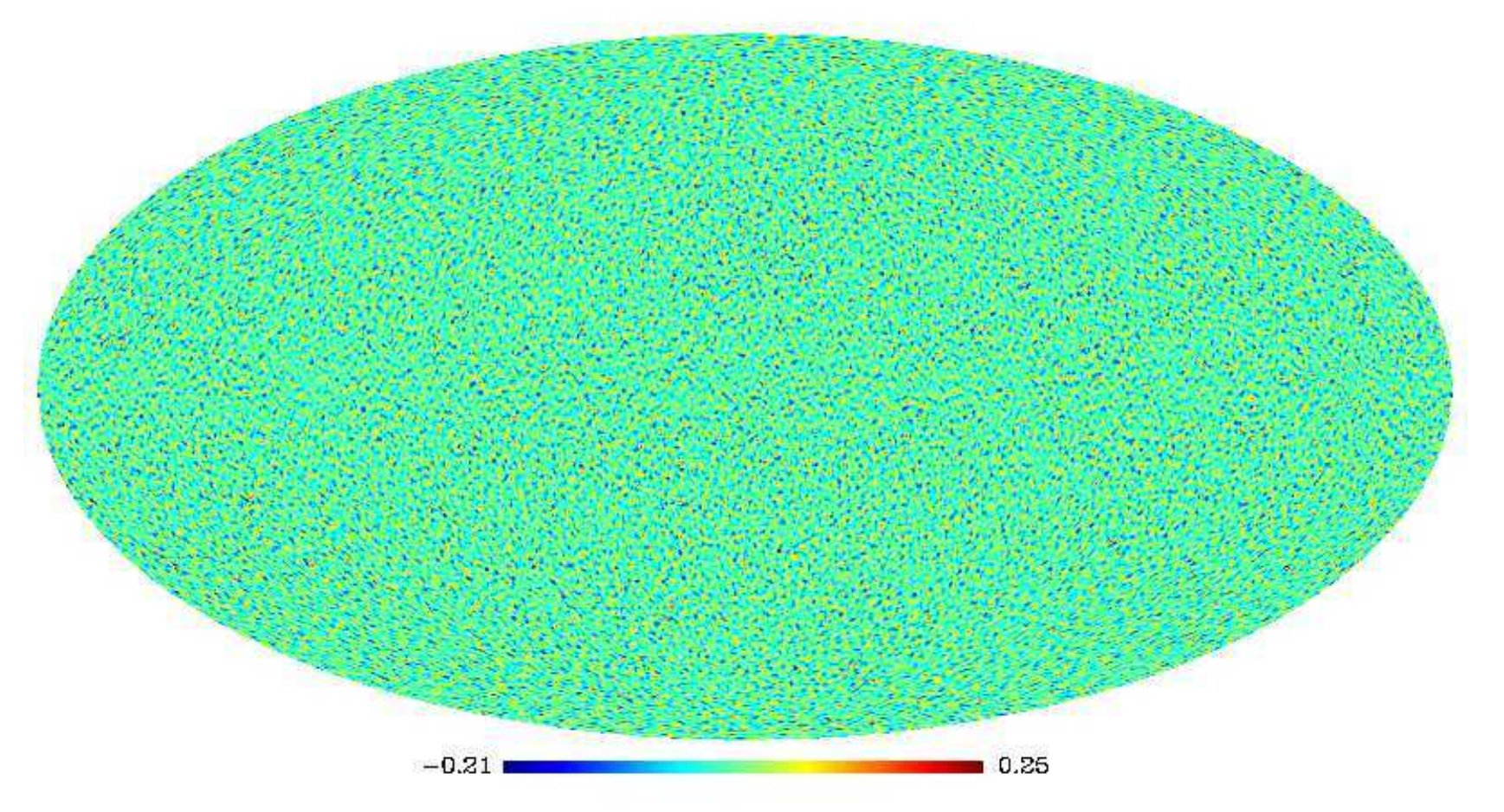}
}
}
\centerline{
\hbox{
\includegraphics[height=4.5cm,width=6.8cm]{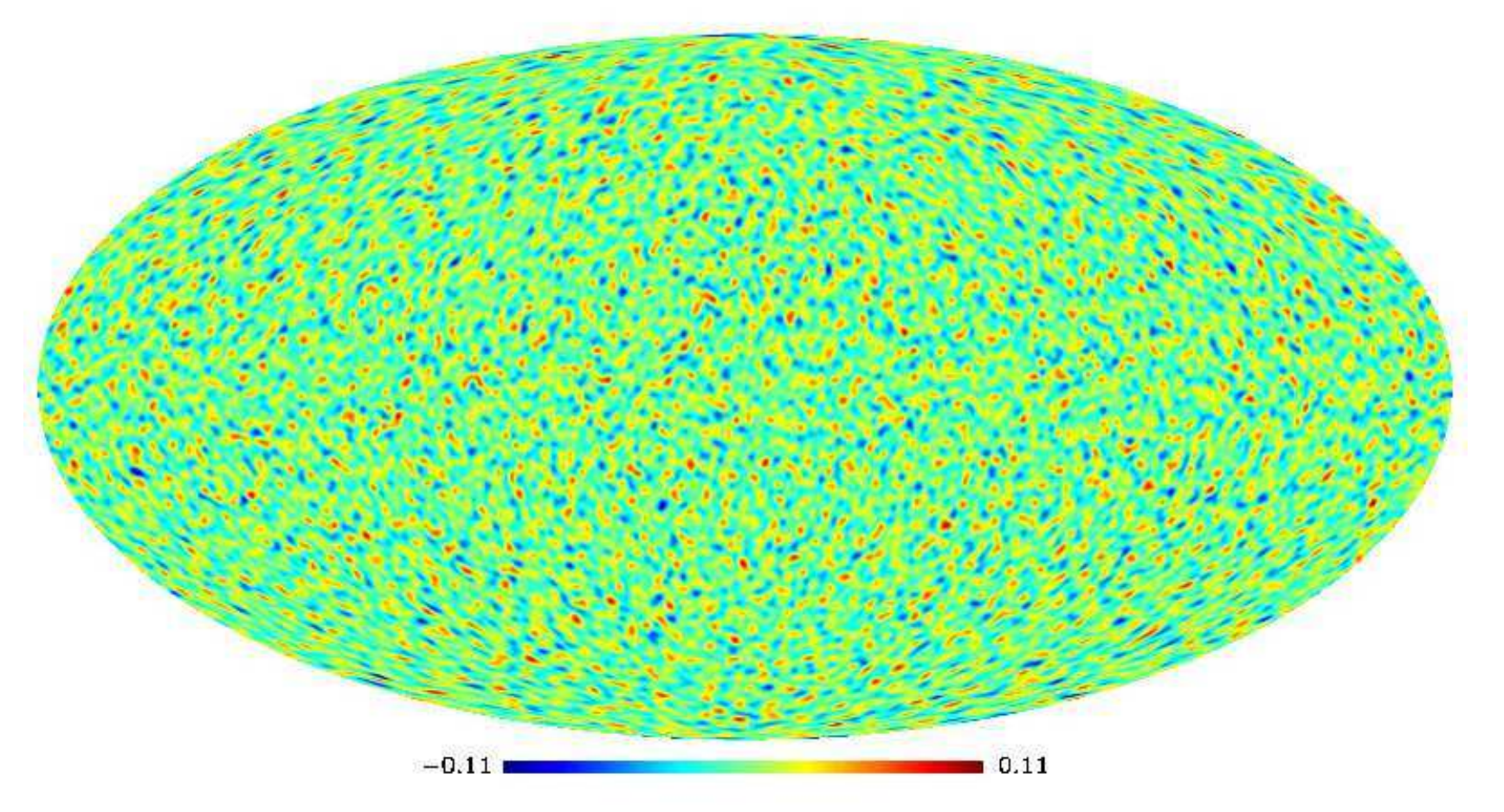}
\includegraphics[height=4.5cm,width=6.8cm]{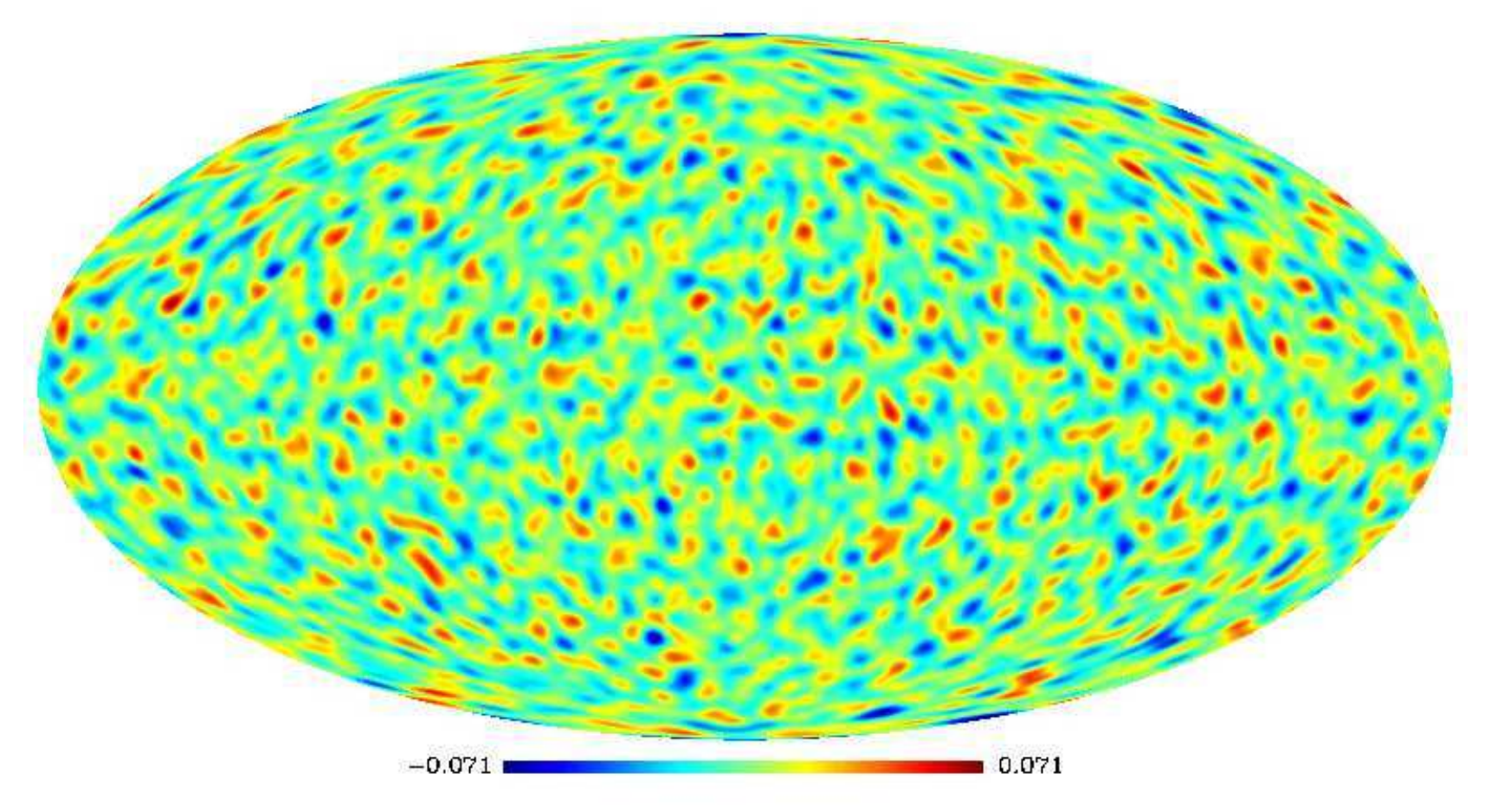}
}
}
\centerline{
\hbox{
\includegraphics[height=4.5cm,width=6.8cm]{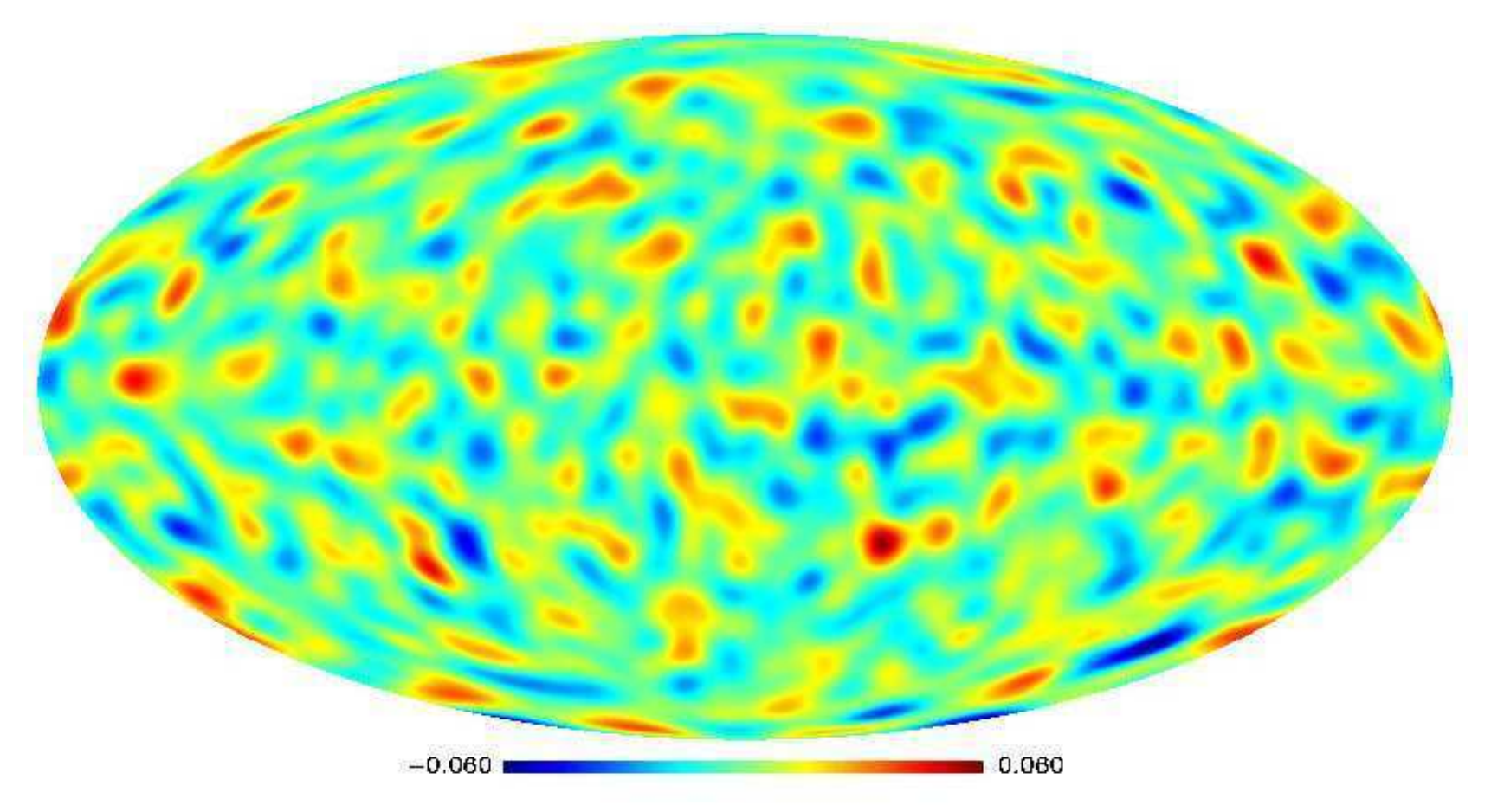}
\includegraphics[height=4.5cm,width=6.8cm]{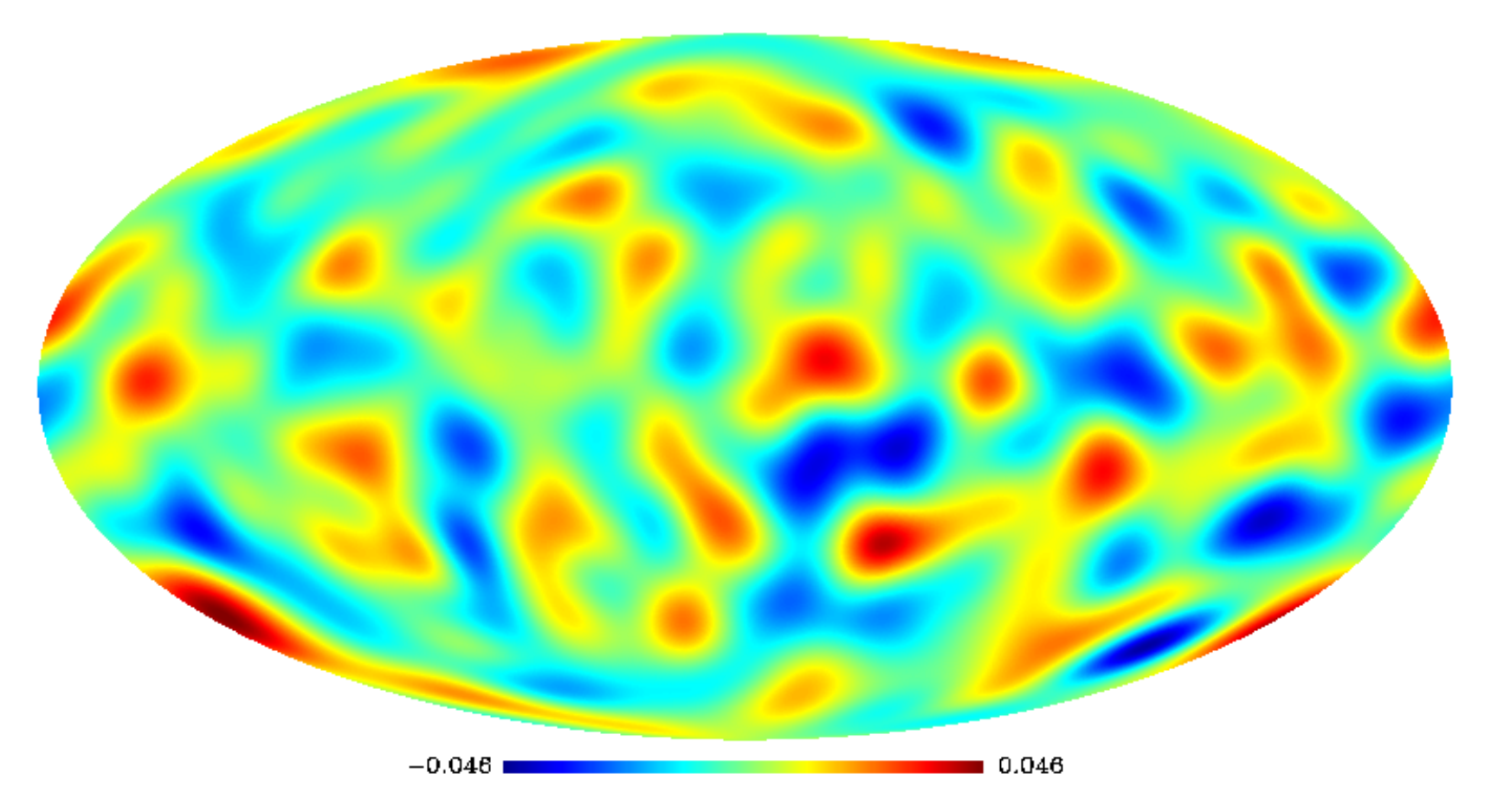}
}
}
\centerline{
\hbox{
\includegraphics[height=4.5cm,width=6.8cm]{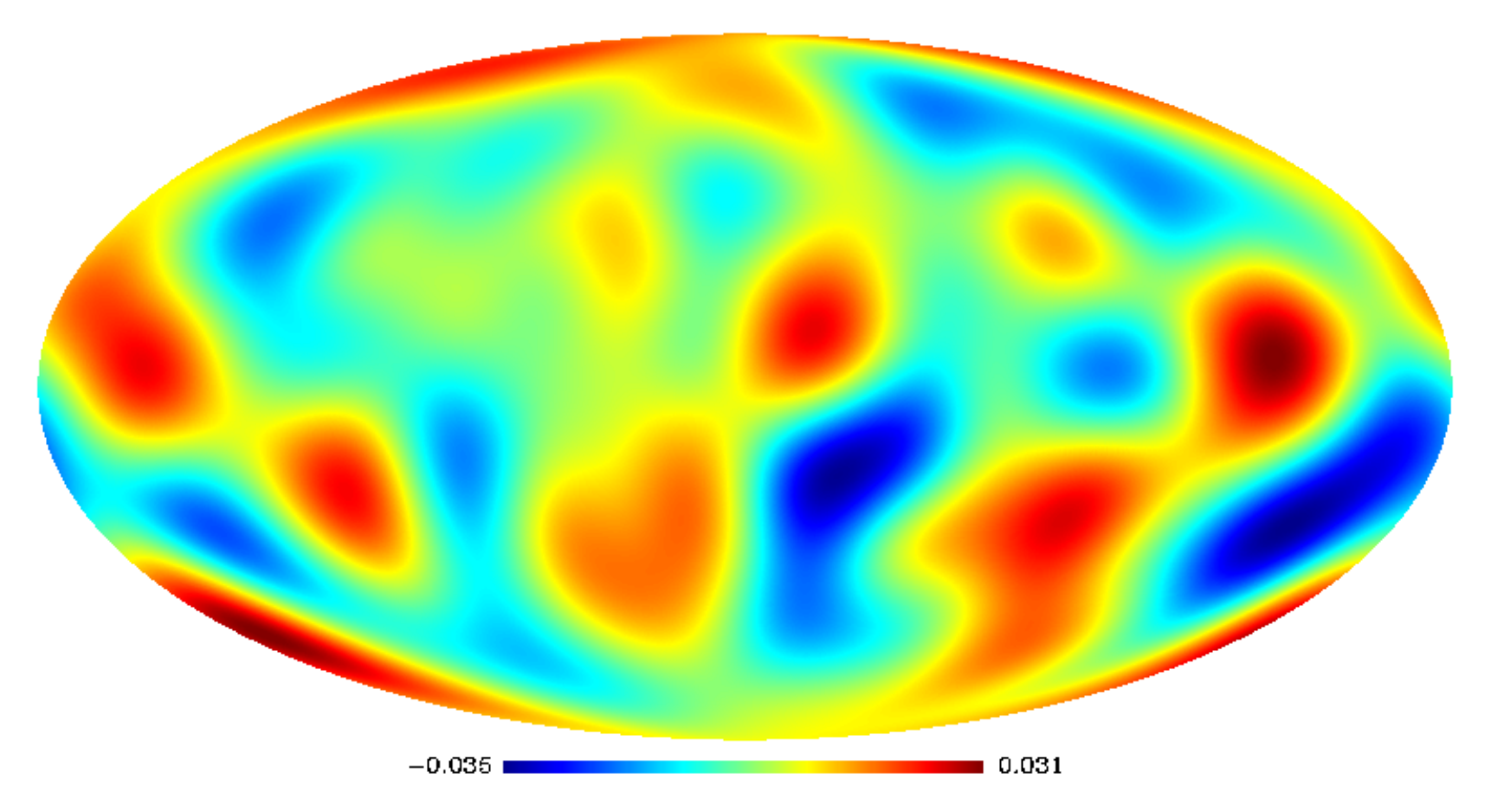}
\includegraphics[height=4.5cm,width=6.8cm]{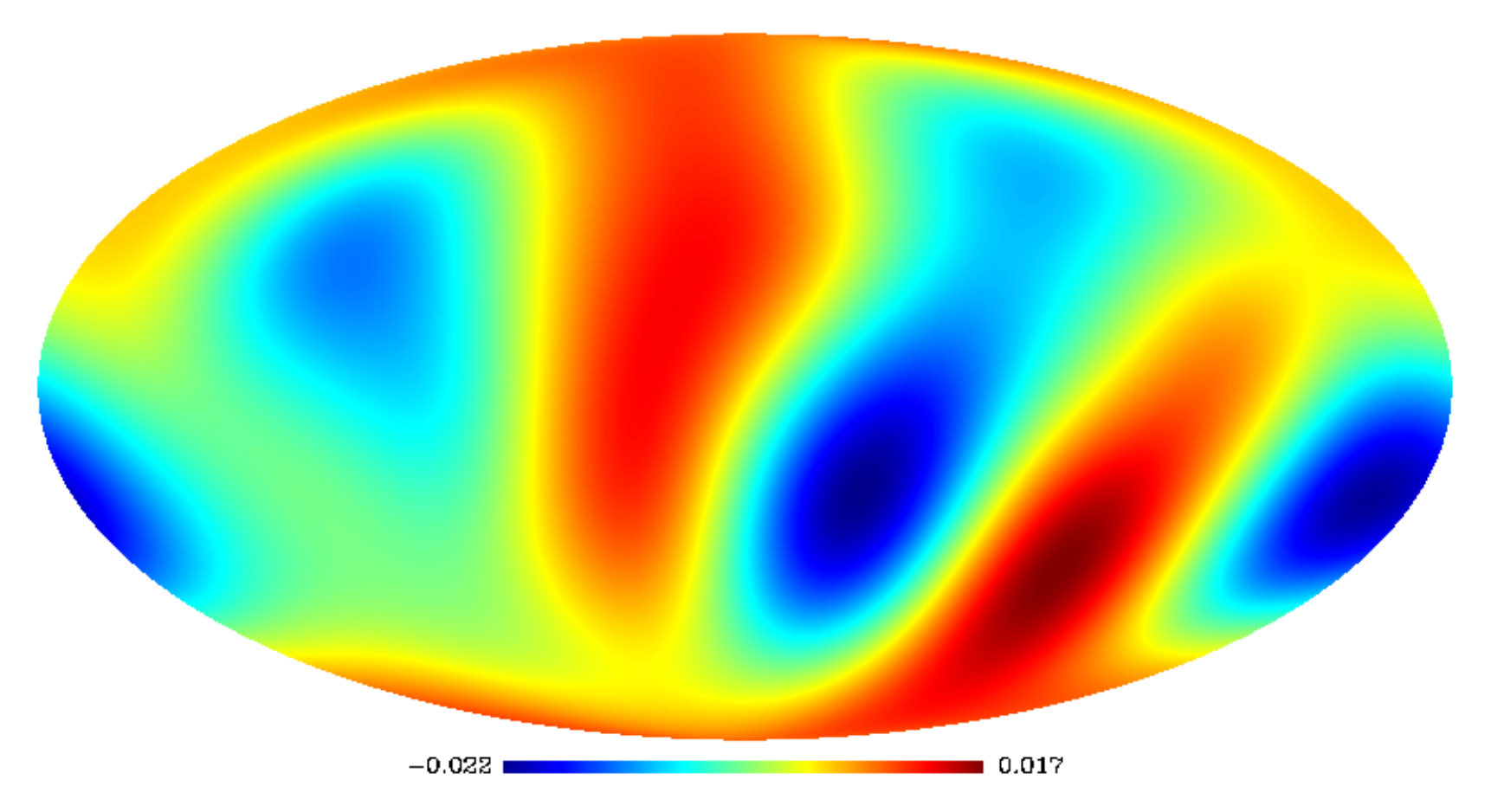}
}
}
}
\caption{\textbf{top left :} masked area. \textbf{from top to bottom and left to right :} the seven wavelet scales of the inpainted map. From the visual point  of view, the initially masked area cannot be distinguished anymore in the wavelet scales of the inpainted map.}
\label{Figure:cmb_scale_wmap_inpainting}
\end{figure}

A simple numerical experiment is shown on figure~\ref{Figure:cmb_wmap_inpainting}. Starting with the full-sky CMB map provided by the WMAP team and available at http://map.gsfc.nasa.gov/. This CMB map was partially masked to discard pixels where the level of  contamination by residual foregrounds is expected to be the highest. Applying the described inpainting algorithm making use of the sparsity of the representation of CMB in the spherical harmonics domain leads to the map shown on the right of figure~\ref{Figure:cmb_wmap_inpainting} : the stationarity of the CMB field appears to have been restored and the masked region is completely undetectable to the eye. Figure~\ref{Figure:cmb_scale_wmap_inpainting} shows the wavelet decomposition of the inpainted map allowing for further visual positive assessment of the quality of the proposed method as again the masked regions are undetectable at all scales.
Power spectrum estimation and non-Gaussianity detection in incomplete CMB maps are more than likely to suffer from the existence of gaps in the data.The numerical experiments on simulated CMB maps we report here tend to show that the proposed inpainting method is able to correctly fill in the missing data thus restoring the \emph{stationarity} of the CMB field and is able to lower the impact of the gaps on the estimation of non-local statistics. The algorithm was applied on nearly a hundred simulated maps which were partly masked using the Wmap's $kp0$ mask, with around 100 iterations in the spherical harmonics representation. Figure~\ref{Figure:powspec} compares the average power spectra estimated from the complete initial maps, to the one retrieved from the masked maps with a sky-coverage correction factor and from the inpainted maps. It appears clearly that inpainting enables a significant reduction of the bias of the spectral estimation compared to the naive coverage correction. However the cost is a slight increase of the variance of the spectral estimator. 
Figure~\ref{Figure:mca_skew_kur} shows plots of the estimated skewness and kurtosis at each scale of the undecimated spherical wavelet transform of both the original map and the inpainted. These statistics are used here as estimators of non-gaussianity. The plots reveal no significant discrepancy: we believe that the proposed method will help discriminate between truly non-gaussian CMB and non-gaussianity related to the non-stationarity of incomplete maps. This will be further investigated in the future.
\section{Conclusion}
\label{conclusion}
This paper presented an inpainting algorithm on the sphere and its application to CMB data analysis. This application relied strongly on the sparse representation of the CMB in the basis of spherical harmonics. The preliminary results shown here allow us to expect that the described inpainting method on the sphere will bear much fruit in the study of CMB, especially concerning CMB power spectrum estimation and in testing for non-Gaussianity in the CMB sky. Thanks to the wealth of  multiscale analysis tools and discrete transforms newly made available for the representation, analysis and synthesis of data on the sphere, we are able to build sparse representations on the sphere of data maps featuring very different structures and morphologies. Hence the proposed inpainting algorithm will easily be applied in other areas where structured data is naturally collected on the sphere.\\

\noindent
\textbf{Acknowledgments : } Some of the results in this paper have been derived using the HEALPix package (G\'orski, Hivon, and Wandelt 1999).

\bibliographystyle{unsrt}
\bibliography{stamet,pixel,eminpaint,gauss,ica,starck,bedros,wave,restore,ima,astro,mc,curvelet,candes,texture,markov,miki,mca,bestbasis}

\end{document}